\documentclass[copyright,creativecommons]{eptcs}
\providecommand{\event}{WFLP 2016}

\usepackage{amssymb}

\renewcommand{\tt}{\ttfamily}
\newcommand{\codefont}{\small\tt}
\newcommand{\code}[1]{\mbox{\codefont{#1}}}
\newcommand{\ccode}[1]{``\code{#1}''}
\newcommand{\fcode}[1]{\mbox{\codefont{\footnotesize{#1}}}} % code in footnote
\newcommand{\mapnd}{\code{\$*}}
\newcommand{\withndarg}{\code{*\$*}}

\usepackage{listings}
\lstnewenvironment{curry}{\lstset{literate={->}{{$\rightarrow{}\!\!\!$}}3}}{}
\newcommand{\listline}{\vrule width0pt depth1.5ex}

\begin{document}

\title{Proving Non-Deterministic Computations in Agda%
\thanks{This material is based upon work partially supported by the National
Science Foundation under Grant No. CCF-1317249
and the German Federal Ministry of Education and Research (BMBF)
under Grant No. 01IH15006B.}
}
\def\titlerunning{Proving Non-Deterministic Computations in Agda}

\author{Sergio Antoy
\institute{Computer Science Dept.\\ Portland State University\\ Oregon, U.S.A.}
\email{antoys@pdx.edu}
\and
Michael Hanus
\institute{Institut f\"ur Informatik\\ CAU Kiel\\ Germany}
\email{mh@informatik.uni-kiel.de}
\and
Steven Libby
\institute{Computer Science Dept.\\ Portland State University\\ Oregon, U.S.A.}
\email{slibby@pdx.edu}
}
\def\authorrunning{S. Antoy, M. Hanus \& S. Libby}

\maketitle

\begin{abstract}
We investigate proving properties of Curry programs
using Agda.
First, we address the functional correctness of Curry functions
that, apart from some syntactic and semantic differences,
are in the intersection of the two languages.
Second, we use Agda to model non-deterministic functions
with two distinct and competitive approaches incorporating the non-determinism.
The first approach eliminates non-determinism by considering the
set of all non-deterministic values produced by an application.
The second approach encodes every non-deterministic
choice that the application could perform.
We consider our initial experiment a success.
Although proving properties of programs is a notoriously difficult task, 
the functional logic paradigm does not seem to add
any significant layer of difficulty or complexity to the task.
\end{abstract}

%------------------------------------------------------------------

\section{Introduction}
\label{Introduction}

The growing interest in software correctness has led to a variety
of approaches for proving, or increasing one's confidence,
that a program computes what is intended.
Examples of this include
\emph{formal verification} \cite{Clarke:1996:FMS:242223.242257},
\emph{design by contract} \cite{Meyer:1992:ADC:618974.619797},
\emph{program analysis} \cite{Nielson:1999:PPA:555142}
and \emph{model checking} \cite{opac-b1084184}.
We are interested in the \emph{functional correctness} of a program,
i.e., the property that each input to the program produces the intended
output.  Agda \cite{Norell:2009:DTP:1481861.1481862}, a language
and system based on \emph{dependent types}, is designed for this purpose.

Agda programs define types and functions. Types are seen as statements,
and functions as constructive proofs of those statements
according to the Curry-Howard isomorphism \cite{curry1980h}.
Agda functions resemble those of Haskell
\cite{PeytonJones03Haskell}.
Agda types extend Haskell's type system since definitions
can be parameterized by (depend on) values of other types.
We are interested in proving properties of Curry programs \cite{Hanus16Curry}.
Curry programs define types and functions that
resemble those of Haskell as well, but with a key difference:
A ``function'' in Curry can be non-deterministic.
Loosely speaking, this means that for the same input, the
function can return one of a set of values.
These \emph{non-deterministic functions}
are not functions in the mathematical sense,
but using them as functions is convenient.
In particular, functional application and functional nesting
provide optimally lazy non-deterministic computations \cite{Antoy97ALP}.

For example, the following Curry code defines a function \code{perm},
that takes a list of elements and
returns any permutation of the list non-deterministically.
\begin{curry}
perm []     = []
perm (x:xs) = ndinsert x (perm xs)
ndinsert x []     = x:[]
ndinsert x (y:ys) = x:y:ys ? y:ndinsert x ys
\end{curry}
The question mark in the last rule denotes the \emph{choice} operation
defined by the rules:
\begin{curry}
x ? y = x
x ? y = y
\end{curry}
The semantics of this definition differs from Agda's and Haskell's.
In the full spirit of declarative programming,
there is no textual order among these rules.
Thus, a value of $t\,\code{?}\,u$ is any value of $t$ or any value of $u$.
The \emph{choice} is substantially equal to McCarthy's
\emph{amb} operator~\cite{McCarthy63}.

Our goal is using Agda to prove the correctness of Curry programs.
This must take into account where Curry differs
from Adga, e.g., non-determinism, the order of evaluation, and non-termination.
Non-determinism may cause some computations to fail, 
and non-termination is essential for Curry's lazy evaluation.
While all these differences are relevant, the crucial one
to handle is non-determinism.

First, we discuss a class of programs that substantially
is the same in Agda and Curry, hence proving the functional
correctness of an Agda program in this class proves it for Curry as well.
Then, we discuss two techniques to cast non-deterministic Curry
programs into programs in the previous class, hence extending
proofs to non-deterministic programs.
The proofs and libraries discussed in this paper are available
at \url{github.com/mihanus/curry-agda}.

\section{Curry}
\label{Curry}

Before discussing our techniques for proving properties of Curry programs,
we briefly review some aspects of functional logic programming, and Curry,
that are necessary to understand the content of this paper.
More details can be found in recent surveys on
functional logic programming \cite{AntoyHanus10CACM,Hanus13}.

Curry \cite{Hanus16Curry} is a declarative, multi-paradigm language,
intended to combine the most important features
of functional and logic programming.
The syntax of Curry is close to Haskell \cite{PeytonJones03Haskell},
only it is extended by allowing
\emph{free} (\emph{logic}) \emph{variables}
in conditions and in right-hand sides of rules.
Moreover, the patterns of a defining rule can be non-linear,
i.e., they might contain multiple occurrences of a variable,
which is an abbreviation for equalities between these occurrences.

The following simple program shows the functional and logic features
of Curry. It defines an operation \ccode{++} to concatenate two lists,
which is identical to the Haskell encoding.
The second operation, \code{ndins}, shows an alternative definition
of the non-deterministic list insertion shown in the
introduction.\footnote{Note that Curry requires the explicit declaration
of free variables, as \fcode{ys} and \fcode{zs} in the rule of \fcode{ndins},
to ensure checkable redundancy.}
\begin{curry}
(++) :: [a] -> [a] -> [a]        ndins :: a -> [a] -> [a]
[]     ++ ys = ys                  ndins x xs | xs == ys$\,$++$\,$zs
(x:xs) ++ ys = x : (xs ++ ys)                 = ys$\,$++$\,$[x]$\,$++$\,$zs   where$\;$ys,zs$\;$free
\end{curry}
Function calls can contain free variables.
They are evaluated lazily where free variables, as demanded arguments,
are non-deterministically instantiated.
Since it is also known that free variables can be replaced
by non-deterministic functions (value generators) \cite{AntoyHanus06ICLP},
we mainly consider non-deterministic functions
instead of free variables in this paper.
Another major difference from Haskell is that
the rules defining an operation are not applied in their textual order.
Any rule which is applicable to some expression is
applied non-deterministically to evaluate this expression.
Thus, if Curry evaluates a call to the choice operation \ccode{?}
defined in the introduction, both rules are applicable.
Hence, the expression \ccode{0$~$?$~$1} evaluates to \code{0} and \code{1}
with the value non-deterministically chosen.

Although non-deterministic functions are a useful programming
feature that leads to new design patterns
\cite{AntoyHanus11WFLP}, they also cause semantic subtleties
that need to be clarified.
For instance, consider the following function to double
the value of an integer number:\label{ex:double}
\begin{curry}
double x = x + x
\end{curry}
One would expect that the result of this operation is always
an even number. Actually, we will show in Sect.~\ref{Examples}
how to prove this property with Agda.
However, it is not obvious that this property holds
with respect to non-deterministic computations.
For instance, one might consider the following derivation:
\begin{curry}
double (0?1)  $\to$  (0?1) + (0?1)  $\to$  0 + (0?1)  $\to$  0 + 1  $\to$  1
\end{curry}
In this derivation, we use two different values for a single argument.
Although one might consider such derivations as acceptable,
which is known as a \emph{run-time choice} semantics,
they come with a potential problem. The result of such
computations depends on the evaluation strategy---an aspect
that should be avoided in a declarative language.
For instance, \code{1} cannot be obtained as a result of
\code{double$\;$(0?1)} with respect to an eager (innermost) evaluation.
Therefore, Curry and similar functional logic languages
adapt the \emph{call-time choice} semantics \cite{Hussmann92}
to obtain a strategy-independent combination of functional
and non-deterministic programming.
Intuitively, an argument passed to a function always has a unique value.
This does not mean that one cannot use non-deterministic functions
as arguments, but it implies that a call like \code{double$\;$(0?1)}
is interpreted like two calls, \code{double$\;$0} and \code{double$\;$1}.
The details of this semantics, known as the CRWL rewriting logic,
can be found in \cite{GonzalezEtAl99}.

\section{Agda}
\label{Agda}

Agda is a dependently typed functional language,
\cite{Bove:2009:BOA:1616077.1616085,Norell:2008:DTP:1813347.1813352}
and is based on intuitionistic type theory.  Similar to Curry, Agda datatypes are
introduced by a data declaration, which gives the name and type of the
datatype as well as the constructors and values in that type.
Agda's types are much more general than Curry's, because 
type definitions may depend on values,
e.g., the type of lists of a given length.
Agda functions are also required to produce a result for every application.
Thus, a function
definition must cover a complete set of patterns and an application
must terminate for every argument.
The parameter passing mode of a function application is by-value,
i.e., the arguments are normalized before the application.
Very loosely speaking, Adga has a richer set of types,
whereas Curry has a richer set of functions.

An Agda program typically defines some types, and
functions over these types.
A later example defines the functions ``\code{+}'', which takes
two natural numbers (in unary representation) and returns their sum,
and \code{even}, which takes a natural number and returns a Boolean value
telling whether the argument is an even number.
A third function, whose identifier is \code{even-x+x},
takes a natural $x$, and has the return type
\code{even$\,$($x$ + $x$) $\equiv$ tt},
where \code{tt} is the Boolean value \emph{true},
\ccode{$\equiv$} is the propositional equality, and ``\code{+}'' and
\code{even} were previously defined.
According to the Curry-Howard isomorphism \cite{curry1980h}, 
the existence of a function of this type, witnessed by its definition,
proves the statement captured by the return type:
for every natural $x$, $x+x$ is even.

Our goal is to prove statements about Curry functions.
Thus, we will define types that capture the statements we wish to prove,
and encode, when we can, definitions of functions of these types.
These will be the proofs of the statements.
An obvious problem is that Curry allows us to encode functions that
cannot be encoded in Agda, and even those that are syntactically similar in the
two languages have significant semantic differences.
Therefore, we will restrict the functions we consider, and discuss
the validity of the statements that we prove despite the semantic differences.

\section{Non-determinism}
\label{Non-determinism}

In this section we discuss two approaches to model non-deterministic
computations in Agda.
Consider the function \code{perm} defined in the introduction.
The Curry evaluation of \code{perm\,[1,2,3]} may return
\code{[2,1,3]}, or any of the six permutations of the
argument.  
Suppose that we wish to prove that for any list $L$,
\code{perm\,$L$} has the same length as $L$.
A crucial task is modeling \code{perm} in Agda, which does not
allow non-determinism.
To understand our approaches, let us first develop a hypothetical
computation of \code{perm\,[1,2,3]} in Curry.  We take some liberties
that will rectify shortly:
\newcommand\xnum[1]{\rlap{\hspace*{23em} \rm (#1)}\mbox{$\to$}}
\begin{curry}
perm [1,2,3] $\xnum 1$ ndinsert 1 (perm [2,3])
             $\xnum 2$ ndinsert 1 [2,3]
             $\xnum 3$ [1,2,3] ? 2 : ndinsert 1 [3]
             $\xnum 4$ 2 : ndinsert 1 [3]
             $\xnum 5$ [2,1,3]
\end{curry}
In line (1) we eagerly evaluate \code{perm\,[2,3]} and arbitrarily
assume that it returns \code{[2,3]}.
In line (3) we arbitrarily choose the right argument of ``\code{?}''.
In line (4) we eagerly evaluate \code{ndinsert\,1\,[3]}, and arbitrarily
assume that it returns \code{[1,3]}.

Since we are going to code \code{perm} in Agda, we must adopt eager
evaluation.  In Curry the order of evaluation does not affect
the result of a computation, as long as the result is obtained.
Thus, for strongly terminating programs,
the functional correctness of a program is independent
of eager vs.{} lazy evaluation.
The real problem is to model the somewhat arbitrary choices made
by non-deterministic functions.

\subsection{Set of values}
\label{Set of values}

Our first approach encodes a Curry non-deterministic function $f$
into an Agda function $f'$, which must be deterministic.
For any
given argument, $f'$ returns the set of all the 
values that $f$ could possibly return.
To this aim, we declare a set-like structure as follows:
\begin{curry}
data ND (A : Set) : Set where
  Val  : A $\to$ ND A
  _??_ : ND A $\to$ ND A $\to$ ND A
\end{curry}
The constructor \code{Val} abstracts a deterministic value in the set.
The constructor ``\code{??}'' makes a set out of two sets, and captures
a non-deterministic choice.

For example, type \code{ND} is used to code an Agda function, \code{perm},
that computes the set of all the permutations of its argument.
\begin{curry}
ndinsert : {A : Set} $\to$ A $\to$ ${\mathbb L}$ A  $\to$ ND (${\mathbb L}$ A)
ndinsert x []        = Val (x :: [])
ndinsert x (y :: ys) = Val (x :: y :: ys)
                    ?? (_::_ y) $\mapnd$ (ndinsert x ys)$\listline$
perm : {A : Set} $\to$ ${\mathbb L}$ A  $\to$ ND (${\mathbb L}$ A)
perm []        = Val []
perm (x :: xs) = ndinsert x *$\code{\$}$* (perm xs)
\end{curry}
The functions \ccode{\mapnd} and \ccode{*\$*} are defined in a
library that we have developed for our
purpose (see Sect.\,\ref{Libraries}).  Function \ccode{\mapnd} is a standard \emph{map} function over
the \code{ND} structure, i.e., it applies the first argument, a
deterministic function, to every value of the second.
Function \ccode{*\$*} is also a \emph{map}, but its first
argument is a non-deterministic function.

With this infrastructure, the statement that any permutation of a list
$L$ has the same length as $L$ is formalized as follows:
\begin{curry}
perm-length : {A : Set} $\to$ (xs : ${\mathbb L}$ A) $\to$
  (perm xs) satisfy ($\lambda$ ys $\to$ length ys =${\mathbb N}$ length xs) $\equiv$ tt
\end{curry}
In the above statement, \code{satisfy} is another library function,
which is infix, with the meaning intended by its name.  The symbols
``${=}{\mathbb N}$'' and ``$\equiv$'' stand for natural number equality and
propositional equality, respectively. The value \code{tt} stands for the Boolean
value \emph{true}.
Informally, the statement is read as ``Given any list \code{xs} of
elements in any set \code{a}, any value \code{ys} of \code{perm xs}
satisfies the condition \code{length ys} $=$ \code{length xs}, where
\code{length} is the usual function that computes the length of a
list.''  The proof of this claim requires a few lemmas, and is approximately
one page long.

\subsection{Planned choices}
\label{Planned choices}

Our second approach encodes a Curry non-deterministic function $f$
into a deterministic Agda function $f'$, that takes an
extra argument.  This argument abstracts the non-deterministic choices
made by $f$ during the computation of a result.  By passing these
choices, $f'$ becomes deterministic and executes the same steps, and
produces the same result, as $f$.

Referring to our non-deterministic computation in the last section, the extra
argument of \code{perm} will encode that, in line (3), the right
argument of ``\code{?}'' has to be selected.  Many proofs of
non-deterministic computations involve statements that hold for
each non-deterministic value, as in our previous example.
Therefore, these statements are independent of the choices made during the computation.
For this reason the extra argument that encodes the choices is abstract,
i.e., never has a concrete value.

With this approach, we parameterize an Agda module with four abstractions:
\begin{enumerate}
\itemsep=0pt
\item \code{Choice\,:\,Set}~~-- a type abstracting non-determinism.
  A value of this type plans the choices that will be made by a
  non-deterministic computation.
\item \code{choose\,:\,Choice $\to {\mathbb B}$}~~--
  a function that returns whether the left or right argument of
  ``\code{?}'' should be selected.
\item \code{lchoice\,:\,Choice $\to$ Choice}~~-- a function that produces
  choices from choices.  The intent is that the choices
  of the result are independent of the choices of the argument.
  The reason for this function will be discussed shortly.
\item \code{rchoice\,:\,Choice $\to$ Choice}~~-- a function like the previous
  one, but it produces other independent choices.
\end{enumerate}
With this machinery, the function that computes a permutation of a list
is defined as follows:
\begin{curry}
ndinsert : {A : Set} $\to$ Choice $\to$ A $\to$ ${\mathbb L}$ A $\to$ ${\mathbb L}$ A
ndinsert _  n []        = n :: []
ndinsert ch n (x :: xs) = if choose ch then n :: x :: xs
                                       else x :: ndinsert (lchoice ch) n xs$\listline$
perm : {A : Set} $\to$ Choice $\to$ ${\mathbb L}$ A  $\to$ ${\mathbb L}$ A
perm _  []        = []
perm ch (y :: ys) = ndinsert (lchoice ch) y (perm (rchoice ch) ys)
\end{curry}
The first required argument of these functions is the plan of the
non-determin\-istic choices.  For example, in the second rule of
\code{ndinsert}, this argument determines whether or not argument
\code{x} should be inserted at the front of the second argument.

The second rule of \code{perm} justifies the necessity of functions
\code{lchoice} and \code{rchoice}.  Both \code{perm} and \code{ndinsert}
make non-deterministic choices.  The choices made by one
function must be independent of the other, otherwise some intended
result could be lost. Hence we need to ``fork'' the choices
encoded by argument \code{ch}.

With this infrastructure, the statement that any permutation of a list
$L$ is as long as $L$ is formalized as follows:
\begin{curry}
perm-length : {A : Set} $\to$ (ch : Choice) $\to$ (xs : ${\mathbb L}$ A)
           $\to$ length (perm ch xs) =${\mathbb N}$ length xs $\equiv$ tt  
\end{curry}
Informally, the statement is read as ``Given any plan of
non-deterministic choices \code{ch}, and any list \code{xs} of elements
in any set \code{a}, the length of the permutation of \code{xs}
according to \code{ch} is the same as the length of \code{xs}.''  The
proof is only a few lines long.

%------------------------------------------------------------------

\section{Libraries}
\label{Libraries}

In order to support the direct translation of Curry programs to Agda,
and to prove theorems about the translated programs,
we developed Agda libraries containing some ubiquitous
functions and theorems.
In particular, the encoding presented in Sect.~\ref{Set of values}
demands such a support. Therefore, we consider only this encoding 
in this section.

Any Curry function can be non-deterministic.
Hence, in our ``set of values'' encoding,
any translated function has a result of type \code{ND}
and must accept arguments of type \code{ND}.
Thus, a Curry function of type
\begin{curry}
$\tau_1~\to~\cdots~\to~\tau_n~\to~\tau$
\end{curry}
should be translated into an Agda function of type
\begin{curry}
$\code{ND}~\tau_1~\to~\cdots~\to~\code{ND}~\tau_n~\to~\code{ND}~\tau$
\end{curry}
Since the type \code{ND} has two constructors,
a direct definition of these Agda functions would be lengthy,
and would also results in lengthy and tedious proofs.
We can simplify the translation based on the following observations:
\begin{itemize}
\item
Some definitional cases are identical for all functions.
An argument of the form \code{$t_1\;$??$\;t_2$}
leads to a non-deterministic result.
The function is applied to both alternatives $t_1$ and $t_2$
and their results are combined with \ccode{??} (this is also called
a ``pull-tab'' step in
\cite{Antoy11ICLP}).
\item
Curry functions are defined with patterns in the left-hand side.
It should be sufficient to define the translated
Agda functions on patterns, and use the \code{ND} type only in the
right-hand side.
\end{itemize}
Therefore, we translate a Curry function of the type shown above
into an Agda function of type
\begin{curry}
$\tau_1~\to~\cdots~\to~\tau_n~\to~\code{ND}~\tau$
\end{curry}
which allows a more direct translation.
We can even improve this translation by analyzing the operation
in more detail. If the function is deterministic, i.e.,
it does not execute non-deterministic choices and calls only deterministic
functions, we can omit the \code{ND} type in the result.
This gives us a one-to-one correspondence between deterministic
Curry functions and their Agda translation.

However, if constructors or functions are applied to other
non-deterministic functions
in the right-hand side of a defining rule, this application is no longer
directly possible, because they do not accept arguments of type \code{ND}.
Therefore, our Agda library provides two operations
to extend such application into a non-deterministic context:
\begin{enumerate}
\item
In order to apply a constructor or deterministic function
to a non-deterministic expression, we define the
application operator \ccode{\mapnd}:
\begin{curry}
_$\mapnd$_ : {A B : Set} $\to$ (A $\to$ B) $\to$ ND A $\to$ ND B
f $\mapnd$ (Val xs)   = Val (f xs)
f $\mapnd$ (t1 ?? t2) = f $\mapnd$ t1 ?? f $\mapnd$ t2
\end{curry}
Thus, \ccode{\mapnd} applies a deterministic function to all values (first rule)
and distributes a non-deter\-ministic choice to
non-deterministic results (second rule).
An application of \ccode{\mapnd} was shown in the
translation of \code{ndinsert} in Sect.~\ref{Set of values}.
\item
To apply a non-deterministic function
to a non-deterministic expression, we define the function \ccode{\withndarg}:
\begin{curry}
_$\withndarg$_ : {A B : Set} $\to$ (A $\to$ ND B) $\to$ ND A $\to$ ND B
f $\withndarg$ (Val x)    = f x
f $\withndarg$ (t1 ?? t2) = f $\withndarg$ t1 ?? f $\withndarg$ t2
\end{curry}
The behavior on choices is similar to \ccode{\mapnd}.
However, for each value \code{x}, the non-deterministic function \code{f}
is applied to obtain a non-deterministic value (first rule).
Therefore, \ccode{\withndarg} is used to combine nested calls
of non-deterministic functions.
The extension to non-deterministic functions
of arities greater than one can be obtained by nested applications
of this operator.
An example application was shown in the
translation of \code{perm} in Sect.~\ref{Set of values}.
\end{enumerate}
In addition to these functions that support the translation
of Curry functions into Agda functions,
our library also includes various operations on the \code{ND} type.
These operations support the formulation of theorems about the translated
Curry programs in Agda.
In general, we are interested in proving properties that are satisfied
by all computed values, like the length property of permutations
shown in Sect.~\ref{Set of values}.
For this purpose, we define a predicate \code{satisfy} (as an infix operator)
which is true if all values in a set satisfy a given predicate:
\begin{curry}
_satisfy_ : {A : Set} $\to$ ND A $\to$ (A $\to$ ${\mathbb B}$) $\to$ ${\mathbb B}$
(Val n)    satisfy p = p n
(t1 ?? t2) satisfy p = t1 satisfy p && t2 satisfy p
\end{curry}
The use of this predicate to formulate a property about the non-deterministic
\code{perm} function was shown in the statement \code{perm-length}
in Sect.~\ref{Set of values}.

When one tries to prove such statements, one often needs to prove
lemmas about properties of the involved functions, like \ccode{\mapnd}
and \ccode{\withndarg}. In order to simplify these proofs,
we also defined and proved a library of Agda theorems to relate
these functions with predicates like \code{satisfy}.
For instance, the following lemma relates \ccode{\mapnd}
and the predicate \code{satisfy} (where $\circ$ denotes function composition,
i.e., \code{(f $\circ$ g) x $\equiv$ f (g x)}):
\begin{curry}
satisfy-$\mapnd$ : {A B$\;$:$\;$Set} $\to$ (f$\;$:$\;$A $\to$ B) (xs$\;$:$\;$ND A) (p$\;$:$\;$B $\to$ ${\mathbb B}$)
           $\to$ (f $\mapnd$ xs) satisfy p $\equiv$ xs satisfy (p $\circ$ f)
\end{curry}
This lemma allows us to eliminate a call to \ccode{\mapnd}
by moving the applied function \code{f} inside the predicate.
The proof of this lemma is easily done by induction on the tree structure
\code{ND}.

The formulation of a similar lemma for \ccode{\withndarg}
is a bit more involved since the applied function is non-deterministic.
However, if we assume the existence of a proof that this function
already satisfies the predicate for all values (given as the
last argument to this lemma),
we can eliminate a call to \ccode{\withndarg}:
\begin{curry}
satisfy-$\withndarg$ : {A B$\;$:$\;$Set} $\to$ (p$\;$:$\;$B $\to$ ${\mathbb B}$) (f$\;$:$\;$A $\to$ ND B) (xs$\;$:$\;$ND A)
           $\to$ ((y$\;$:$\;$A) $\to$ (f y) satisfy p $\equiv$ tt) $\to$ (f $\withndarg$ xs) satisfy p $\equiv$ tt
\end{curry}
Our library contains more predicates and lemmas that are useful
to state and prove specific properties of Curry programs.
For instance, to show that some non-deterministic Boolean expression
is always true, one can use the following predicate:
\begin{curry}
always : ND ${\mathbb B}$ $\to$ ${\mathbb B}$
always (Val b)    = b
always (t1 ?? t2) = always t1 && always t2
\end{curry}
A lemma to eliminate an occurrence of \ccode{\mapnd}
in the context of \code{always} is the following:
\begin{curry}
always-$\mapnd$ : {A$\;$:$\;$Set} $\to$ (p$\;$:$\;$A $\to$ ${\mathbb B}$) (xs$\;$:$\;$ND A)
          $\to$ ((y$\;$:$\;$A) $\to$ p y $\equiv$ tt) $\to$ always (p $\mapnd$ xs) $\equiv$ tt
\end{curry}
It states that, if there is a proof that \code{p} is always true,
the application of \code{p} to a non-deterministic value
is also true.
An application of this lemma is shown in the following section,
where we discuss some examples to prove properties
of Curry programs.

%------------------------------------------------------------------

\section{Examples}
\label{Examples}

\subsection{Example 1}

In this example, we assume that natural numbers are defined
in Peano representation as follows:
\begin{curry}
data Nat = Z | S Nat
\end{curry}
The following Curry function non-deterministically yields
an even and an odd number close to its input value:
\begin{curry}
eo :: Nat -> Nat
eo n = n ? (S n)
\end{curry}
To show some interesting property related to \code{eo},
consider the function \code{double}, shown in Sect.~\ref{ex:double},
and the predicate \code{even}:
\begin{curry}
even :: Nat -> Bool
even Z         = True
even (S Z)     = False
even (S (S x)) = even x
\end{curry}
We want to prove that, for all natural numbers $n$,
\code{even$\;$(double$\;$(eo $n$))} is always \code{True}.
Note that this property holds only for the call-time choice semantics
of Curry (see Sect.~\ref{Curry}).
In the run-time choice semantics, or term rewriting, this property does
not hold.

To prove this property with Agda, we have to translate
the program to Agda.
Since \code{double} and \code{even} are deterministic functions,
they can be directly translated into Agda functions, where
the constructors of the Curry program are translated into
corresponding constructors already defined in Agda:
\begin{curry}
double : ${\mathbb N}$ $\to$ ${\mathbb N}$              even : ${\mathbb N}$ $\to$ ${\mathbb B}$
double x = x + x         even zero          = tt
                         even (suc 0)       = ff
                         even (suc (suc x)) = even x
\end{curry}
Since non-determinism is not involved here,
we can prove that \code{even$\;$(double $n$)} is always true
by standard techniques. For this purpose, we state the lemma
\begin{curry}
even-x+x : (x : ${\mathbb N}$) $\to$ even (x + x) $\equiv$ tt
\end{curry}
which is immediately proved by induction on its argument.
Exploiting this lemma, we can prove our intended property:
\begin{curry}
even-double-true : (x : ${\mathbb N}$) $\to$ even (double x) $\equiv$ tt
even-double-true x rewrite even-x+x x = refl
\end{curry}
In order to prove our initial property about the expression
\code{even$\;$(double$\;$(eo $n$))}, we have to decide about the
representation of non-determinism introduced in
Sect.~\ref{Non-determinism}.
If we choose the ``set of values'' representation,
the function \code{eo} is defined as follows:
\begin{curry}
eo : ${\mathbb N}$ $\to$ ND ${\mathbb N}$
eo n = Val n ?? Val (suc n)
\end{curry}
Using the library functions presented in the previous section,
the Curry expression \code{even$\;$(double$\;$(eo$\;n$))}
is represented in Agda as:
\begin{curry}
(even $\circ$ double) $\mapnd$ (eo n)
\end{curry}
The proof of our property is just an application
of the lemma \code{always-\mapnd} (see Sect.~\ref{Libraries})
to the previous lemma:
\begin{curry}
even-double-eo-true : (n : ${\mathbb N}$) $\to$ always ($\mapnd$ (even $\circ$ double) (eo n)) $\equiv$ tt
even-double-eo-true n = always-$\mapnd$ (even $\circ$ double) (eo n) even-double-true
\end{curry}
The second ``choice'' representation of non-determinism
provides for a more direct proof.
We pass the plan of non-deterministic choices as the first argument to
\code{eo}:
\begin{curry}
eo : Choice $\to$ ${\mathbb N}$ $\to$ ${\mathbb N}$
eo n = if choose ch then n else (suc n)
\end{curry}
Then the proof is an immediate application of the lemma
\code{even-double-true}:
\begin{curry}
even-double-eo-true : (ch: Choice) (n : ${\mathbb N}$) $\to$ even (double (eo ch n)) $\equiv$ tt
even-double-eo-true ch n = even-double-true (eo ch n)
\end{curry}
As we can see, the proof with the ``choice'' representation
is more direct, and shorter, than with the ``set of values''
representation, as already mentioned in Sect.~\ref{Non-determinism}
for the property of the length of permutations.

\subsection{Example 2}

So far we have proved properties about predicates,
however, we have not discussed any notion of equality.
In Agda, two terms are equal if their values are syntactically equal.
This is denoted by the standard definition:
\begin{curry}
data _$\equiv$_ {A : Set} (x : A) : A $\to$ Set where
  refl : x $\equiv$ x
\end{curry}
e.g., \code{1+1}$\;\equiv\;$\code{2}.

However, we cannot apply this relation to non-deterministic values
for the simple reason that deterministic
values and non-deterministic values are structurally different.
For example, given the non-deterministic value:
\begin{curry}
coin = (Val tt) ?? (Val ff)
\end{curry}
it would seem reasonable to state that
\code{tt $\equiv$ coin},
but \code{coin} reduces to \code{(Val tt)$\;$??$\;$(Val ff)}
which is not the same value as \code{tt}.

Instead, we must define a new notion of equality for non-deterministic values.
We chose an existential interpretation of equality for our representation.
If a value $x$ exists inside the set of values of a
non-deterministic expression $nx$, then the two are considered equivalent
(non-deterministically equal).
A proof of equivalence is then reduced to finding a path through the tree representing $nx$.
This leads to the following structure for proofs of equivalence.
\begin{curry}
data _$\in$_ {A : Set} (x : A) : (nx : ND A) $\to$ Set where
  ndrefl : x $\in$ (Val x)
  left   : (l : ND A) $\to$ (r : ND A) $\to$ x $\in$ l $\to$ x $\in$ (l ?? r)
  right  : (l : ND A) $\to$ (r : ND A) $\to$ x $\in$ r $\to$ x $\in$ (l ?? r)
\end{curry}
Now we can give a proof that \code{tt} is in \code{coin}

\begin{curry}
hInCoin : tt $\in$ coin
hInCoin = left (Val tt) (Val ff) ndrefl
\end{curry}
This can be read as \code{tt} is on the left side of the tree
\code{(Val tt)$\;$??$\;$(Val ff)}.

Now that we have a notion of equivalence,
we can prove some more interesting results.
Recall that \ccode{\mapnd} applies a function \code{f}
to every value of a non-deterministic expression $nx$.
For consistency, it would be good to prove that if \code{x $\in$ nx}
then \code{f x $\in$ f $\mapnd$ nx}.
This is a simple structural induction.

\begin{curry}
$\in$-$\mapnd$ : {A B : Set} $\to$ (f : A $\to$ B) $\to$ (x : A) $\to$ (nx : ND A)
     $\to$ x $\in$ nx $\to$ f x $\in$ f $\mapnd$ nx$\listline$
$\in$-$\mapnd$ f x (Val .x) ndrefl = ndrefl
$\in$-$\mapnd$ f x (l ?? r) (left$~$ .l .r k) = left$~$ (f $\mapnd$ l) (f $\mapnd$ r) ($\in$-$\mapnd$ f x l k)
$\in$-$\mapnd$ f x (l ?? r) (right .l .r k) = right (f $\mapnd$ l) (f $\mapnd$ r) ($\in$-$\mapnd$ f x r k)
\end{curry}
We can prove a similar result for \ccode{\withndarg}.
The proof is nearly identical to \code{$\in$-$\mapnd$},
but it has the general type:
\begin{curry}
$\in$-$\withndarg$ : {A B : Set} $\to$ (x : A) $\to$ (nx : ND A) $\to$ (f : A $\to$ B)
      $\to$ (nf : A $\to$ ND B) $\to$ x $\in$ nx $\to$ f x $\in$ nf x $\to$ f x $\in$ nf $\withndarg$ nx
\end{curry}
For a more interesting example, consider the problem of sorting a list.
The problem of determining if a list is sorted has been studied extensively,
but we would like to look at a conceptually more difficult problem.
How can we determine if a sorted list is actually a permutation of the original list?

This has also been studied before now,
and the solutions tend to be long and difficult to follow.
A search of ``github,'' ``agda'' and ``sort'' gives
half a dozen independent efforts.
Since we defined the non-deterministic permutation function above,
it would be nice to have a theorem like (for the sake of simplicity,
we consider sorting natural numbers):
\begin{curry}
sortPerm : (xs : ${\mathbb L}$ ${\mathbb N}$) $\to$ sort xs $\in$ perm xs
\end{curry}
This is short, simple, and obviously correct if we trust \code{perm} to produce only permutations of a list.
For this proof we will be using the following simple insertion sort.

\begin{curry}
insert : ${\mathbb N}$ $\to$ ${\mathbb L}$ ${\mathbb N}$ $\to$ ${\mathbb L}$ ${\mathbb N}$
insert x []        = x :: []
insert x (y :: ys) = if x < y then (x :: y :: ys)
                              else (y :: insert x ys)$\listline$
sort : ${\mathbb L}$ ${\mathbb N}$ $\to$ ${\mathbb L}$ ${\mathbb N}$
sort []        = []
sort (x :: xs) = insert x (sort xs)
\end{curry}
We only need two lemmas.
The first states that non-deterministic equivalence holds over conditional statements.
This result allows us to introduce conditional
statements when talking about equivalence.

\begin{curry}
ifIntro : {A : Set} $\to$ (x : A) $\to$ (y : A) $\to$ (nx : ND A)
        $\to$ x $\in$ nx $\to$ y $\in$ nx $\to$ (c : ${\mathbb B}$) $\to$ (if c then x else y) $\in$ nx
\end{curry}
The second lemma is an equivalence between our \code{insert} function
and the non-deterministic \code{ndinsert} defined above.
This will give the intuitive result that, for any list \code{xs},
inserting a new element
does not violate our non-deterministic equivalence.
The type of this lemma is given by: 
\begin{curry}
insert=ndinsert : (y : ${\mathbb N}$) $\to$ (xs : ${\mathbb L}$ ${\mathbb N}$) $\to$ insert y xs $\in$ ndinsert y xs
\end{curry}
The proof of this lemma contains two trivial cases, but the final
case gives some insight into working with non-deterministic values.
The idea is straightforward.  Either the value \code{y}
is inserted at the front of the list, or it is inserted somewhere in the
tail.  Since our non-deterministic \code{ndinsert} covers both of those cases
we do not need to know exactly where it is inserted.  We just pick the
correct case.

The full code is given below for completeness.
Variables \code{l} and \code{r} represent the left and right branches in \code{insert} respectively,
while \code{nl} and \code{nr} are the corresponding branches for \code{ndinsert}.
Additionally, \code{rec} is a recursive case if
\code{y} is inserted in the tail of the list,
Finally \code{l$\in$step} and \code{r$\in$step} are proofs that \code{l} and \code{r} are somewhere in the result of inserting
\code{y} into the list.

\begin{curry}
insert=ndinsert y (x :: xs) = ifIntro l r step l$\in$step r$\in$step (y < x) where
  step   = ndinsert y (x :: xs)
  l      = (y :: x :: xs)
  r      = x :: insert y xs
  nl     = Val (y :: x :: xs)
  nr     = (_::_ x) $\mapnd$ (ndinsert y xs)
  rec    = $\in$-$\mapnd$ (_::_ x) (insert y xs) (ndinsert y xs) (insert=ndinsert y xs)
  l$\in$step = left  nl nr ndrefl
  r$\in$step = right nl nr rec
\end{curry}
Finally, we have the tools to prove our sorting theorem.
After the previous result the theorem is remarkably simple.

\begin{curry}
sortPerm : (xs : ${\mathbb L}$ ${\mathbb N}$) $\to$ sort xs $\in$ perm xs
sortPerm [] = ndrefl
sortPerm (x :: xs) = $\in$-$\withndarg$ (sort xs) (perm xs) (insert x) (ndinsert x) 
                           (sortPerm xs) (insert=ndinsert x (sort xs))
\end{curry}

%------------------------------------------------------------------

\section{Partial Functions}
\label{Partial Functions}

Up to now we have considered non-deterministic but totally defined functions
in our examples.
However, functions in Curry can also be partially defined,
and failures caused by such partial definitions are not a run-time
error as in purely functional programming but a desirable feature
that frees the programmer from specifying all computation details
in a program.
For instance, consider the problem of selecting a minimum
element in a list of numbers. Such a function is easily
defined in Curry:
\begin{curry}
minND :: [Int] -> Int
minND xs@(_ ++ [x] ++ _) | all (x<=) xs = x
\end{curry}
To specify the selection of an arbitrary element
without a concrete strategy,
we use a functional pattern \cite{AntoyHanus05LOPSTR}
in the left-hand side of \code{minND}.
The condition restricts the application of the rule
to all selected elements that are less than or equal to all other elements.
Thus, the condition fails on elements that are not minimal.

Although computations with failing branches are an important
feature of functional logic languages,
the modeling of failures in Agda is not straightforward.
First of all, Agda requires that all functions are totally
defined. Typically, the application of partially defined functions, like
\code{head} or \code{tail} on lists, is restricted
by a refined type, e.g., by requiring a proof that
the argument of \code{head} or \code{tail} is a non-empty list.
In our case this is not a reasonable solution
since a Curry program actually deals with failed computations.

Another alternative to deal with partially defined functions
is to extend the result type to indicate a non-successful
application of a function. For instance, one can use
the \code{Maybe} type to define \code{head} as a total function in Agda:
\begin{curry}
head : {A : Set} $\to$ ${\mathbb L}$ A $\to$ maybe A
head (x :: _) = just x
head []       = nothing
\end{curry}
Unfortunately, this modeling is limited to ``top-level'' failures.
For instance, doing the same for \code{tail} would inhibit
nested applications of \code{tail} due to typing reasons.

Another, and more serious, problem is the fact that
Curry, as a lazy language, can also ignore failures
if they occur in arguments that are not demanded.
For instance, the Curry expression
\begin{curry}
head (0 : tail [])
\end{curry}
evaluates to \code{0} since the failure, which occurs if \code{tail$\;$[]}
is evaluated, is ignored thanks to laziness.
On the other hand, in Agda one cannot construct a list containing
failures in the tail. To implement a similar expression
in Agda, one has to evaluate, before constructing the list,
the tail \code{tail$\;$[]}, which returns \code{nothing},
so that the entire computation fails.
This has the consequence that, if we prove in Agda some property
about \emph{all} successful computations, it might not be transferable
to Curry, because Curry might produce additional successful computations.

A possible solution to this problem could be a different
modeling of Curry's data structures in Agda.
For instance, instead of mapping Curry lists to Agda lists,
one could include failures and non-deterministic choices
as list constructors:
\begin{curry}
data List (A : Set) : Set where
  empty     : List A
  cons      : A $\to$ List A $\to$ List A
  Fail-List : List A
  _??-List_ : List A $\to$ List A $\to$ List A
\end{curry}
Then one can completely define \code{tail} on this structure:
\begin{curry}
tail : {A : Set } $\to$ List A $\to$ List A
tail (cons _ xs)     = xs
tail empty           = Fail-List
tail Fail-List       = Fail-List
tail (l1 ??-List l2) = tail l1 ??-List tail l2
\end{curry}
With such a representation, we can ``totalize''
all functions, and allow failures inside data structures.
Basically, it is quite similar to the translation
of Curry programs into Haskell, which is the basis
of the Curry implementation KiCS2 \cite{BrasselHanusPeemoellerReck11}.
However, it is also known that this translation models
the run-time choice semantics instead of the call-time choice semantics
used in Curry (see Sect.~\ref{Curry}).
To obtain a call-time choice semantics, one can attach
unique identifiers to each choice constructor
in order to make consistent selections of computed values
(see \cite{BrasselHanusPeemoellerReck11} for details).
Although this is a systematic method to translate
Curry programs into purely functional programs,
it does not seem adequate for Agda since the generation
of unique identifiers requires infinite data structures,
and the resulting code is quite complex, requiring more
complex proofs.

Although the handling of partially defined functions
is difficult in general,
one can still use Agda to prove properties of such functions,
if the failures occur at the computation's top-level
so that one can reason about their occurrences.
For instance, consider the well-known functional logic
definition of computing the last element of a list by exploiting
the list concatenation \ccode{++}:
\begin{curry}
last :: [a] -> a
last xs | ys ++ [x] == xs  = x   where ys,x free
\end{curry}
We want to show that \code{last} is a deterministic operation,
i.e., for a fixed input list $xs$, the evaluation of \code{last$\;xs$}
cannot yield two different results.
Although the definition of \code{last} is partial, and uses
free variables and non-determinism, the formulation and proof
of this property is not difficult in Agda by considering
only the condition of \code{last}.
We show that, if there are two pairs of values for \code{ys} and \code{x}
that satisfy the condition for some fixed \code{xs},
the two values for \code{x} are identical.
This property can be expressed (with some type specialization to increase
readability) in Agda as follows:
\begin{curry}
last-det : (ys1 ys2 : ${\mathbb L}$ ${\mathbb N}$) (x1 x2 : ${\mathbb N}$)
         $\to$ =${\mathbb L}$ _=${\mathbb N}$_ (ys1 ++ [ x1 ]) (ys2 ++ [ x2 ]) $\equiv$ tt  $\to$  x1 =${\mathbb N}$ x2 $\equiv$ tt
\end{curry}
As a further example, consider again the non-deterministic and partial
function \code{minND} from the beginning of this section.
We want to implement the same functionality in a deterministic manner
by a list traversal:
\begin{curry}
minD :: [Int] -> Int
minD [x]      = x
minD (x:y:xs) = let z = minD (y:xs) in if x <= z then x else z
\end{curry}
Since the correctness of this list traversal is not obvious,
we take the definition of \code{minND} as the specification
and prove that \code{minD} satisfies the specification,
a typical task in (declarative) program development
\cite{AntoyHanus12PADL}.
If we model \code{minND} with the ``planned choices'' translation,
this property could be formulated in Agda as follows
(the additional argument in the Agda function \code{minD}
is a proof that the argument list is not empty; this is required
since \code{minD} is partially defined but Agda accepts only total
functions):
\begin{curry}
minlist-theorem : (ch : Choice) (x : ${\mathbb N}$) (xs : ${\mathbb L}$ ${\mathbb N}$) (z : ${\mathbb N}$)
             $\to$ minND ch (x :: xs) $\equiv$ just z $\to$ z =${\mathbb N}$ minD (x :: xs) refl $\equiv$ tt
\end{curry}
We use a \code{maybe} result type to express the partiality of
the non-deterministic function \code{minND} and a proof
of non-emptiness for the partiality of the deterministic function
\code{minD} (as discussed above).
Then the theorem states that, if \code{z} is a minimal value
according to the specification, it will be computed by
the deterministic implementation.
The proof is non-trivial and requires various lemmas.
In particular, the proof is split into showing that the
non-deterministically selected element is smaller than, or equal to, than
all other list elements, but it cannot be strictly smaller than all
others since it is itself an element from the list.

%------------------------------------------------------------------

\section{Semantics}
\label{Semantics}

Since we prove properties of Agda functions, but are interested
in properties of Curry functions, we discuss how to migrate
from one environment to the other.

Initially, we consider
datatypes, such as algebraically defined natural numbers,
lists and trees, and functions operating on these types
that have equal definitions, except for minor syntactic differences,
in the two languages.  We call the set of such types and functions
the \emph{intersection} of the two languages.
For example, a function such as ``length of a list'',
returning a natural number, is in the intersection.
Agda functions, including those in the intersection,
are totally defined, terminating, and evaluated eagerly.
The rules defining these functions are
unconditional, deterministic, and without free variables.
Any expression in the intersection has a value, because it is an Agda
expression, and this value is the same in the two languages,
because in this situation eager and lazy evaluations produce
the same result.  Thus, our first claim is that
any statement about the input/output behavior of
a function, i.e., its functional correctness, proved in Agda
holds for Curry.

Then, we define Agda functions that simulate non-deterministic
Curry functions.  For example, a function such as
``permutation of a list'' simulates the corresponding
non-deterministic Curry function.  Regardless of the approach,
either \emph{set of values} or \emph{planned choices},
these Agda functions are still in the intersection
of the two languages.  Thus, we can compare a simulated
non-deterministic function (in Curry and Agda the behavior
is the same) against the same native non-deterministic
Curry function.  Of course, the expectation is that the
simulated and the native functions have the same behavior.
Proving this fact would not involve Agda.

Curry has a rich syntax, and some high-level constructs that are
convenient for programming, and are not in the intersection
with Agda.  These constructs can sometimes be mapped to more
elementary Curry constructs that are in the intersection
of the two languages,
hence the set of Curry functions that we can reason about becomes larger.
For example, Curry allows free variables in expressions that are instantiated
by narrowing \cite{Reddy85,Slagle74}.
Free variables are replaced by generators \cite{AntoyHanus06ICLP},
non-deterministic functions, that are the focus of our work.
Generators are typically non-terminating functions, shortly we will discuss
a workaround.
Curry also allows conditional rules. With some care,
the conditions of conditional rules are moved
to the right-hand sides \cite{Antoy01PPDP}.
Care is needed since Agda disallows incompletely defined functions.
When there are conditions it may be difficult, or
impossible, to tell whether a function is defined for all inputs.
Finally, Curry allows functions that do not terminate because
of infinite recursion.
When one such function is applied in a computation,
the lazy evaluation strategy makes a recursive call only
when the result of the call is demanded.
To avoid non-termination in Agda, we add an additional argument
that limits the number of recursive calls.

%------------------------------------------------------------------

\section{Related Work}
\label{Related Work}

There are only a few works on verifying properties of
functional logic programs. One reason may be the fact
that declarative programs are often considered as clear
specifications so that there is nothing to prove about them.
However, if one develops a more efficient version
of a declarative program, because the initial version is not
efficient enough in a particular application context,
one wants to prove that the new version behaves like the
initial version. This has been formalized in \cite{AntoyHanus12PADL},
where the notions of specifications and contracts
for functional logic programs have been introduced
together with some examples and proof obligations.
Concrete methods to prove such obligations were not discussed.
We have shown in this paper that it is possible to use Agda
for this purpose, e.g., by proving that insertion sort computes
a permutation of the input list, or that a non-deterministic
selection of a minimum of a list is equivalent to a deterministic
selection algorithm.

The use of proof assistants to verify functional logic programs
has been pioneered in \cite{ClevaLeachLopez04PPDP},
where the call-time choice semantics has been
formalized with different proof tools. Although the authors
formally justified their translation by mapping
the CRWL rewriting logic \cite{GonzalezEtAl99} into logic programs,
only trivial properties, like \code{0 $\in$ double$\;$(0?1)},
are proved.
A similar approach is used in \cite{ClevaPita06} where
CRWL is translated into the specification language
Maude, and some non-trivial but (compared to our approach) still
simple properties are proved.

Since the formalization of the call-time choice semantics
causes complex proofs, the proof method has been improved
in \cite{ClevaLopezFraguas07} by a characterization
of deterministic functions and expressions, and by using
specific proof techniques for them. Although the authors
were able to construct short proofs for deterministic properties
inside functional logic programs,
like the commutativity of addition, proofs for more complex properties
have not been reported.
These CRWL-translation based approaches can deal
with the full class of functional logic programs, whereas
our limited framework is able to verify non-trivial properties.

%------------------------------------------------------------------

\section{Conclusion}
\label{Conclusion}

We have used the Agda system to verify the functional correctness of
Curry functions.  Our focus has been on Curry's non-deterministic
functions, a device akin to a function that returns one of many values
for the same input.  Such a device contributes the logic component
of the language in a lazy functional setting.

We modeled non-determinism in two different ways.
One considers at once all the non-deterministic values
that a function may return.
The other adds an argument to a function that encodes
every non-deterministic choice that the function may make.
Both models eliminate the non-determinism of a functions
and enable us to use Agda, which is deterministic, to
prove properties of non-deterministic functions.

We developed some libraries that define non-deterministic structures
and prove generic lemmas about them.
With the help of these libraries, we proved some interesting
properties of Curry programs.
We highlighted the semantic differences between Agda and Curry, and
discussed why, despite these differences, theorems proved in Agda
migrate to Curry.

Proving properties of programs is a difficult task.
Our effort shows that proving properties of Curry programs
with Adga is possible, and that
the logic component of Curry is not a major obstacle.

%------------------------------------------------------------------

\bibliographystyle{eptcs}
\bibliography{paper}

\end{document}